\documentclass[a4paper]{jpconf}
\usepackage{graphicx,amsmath,amssymb,bm,mathrsfs}

\newcommand*{\ddpar}[2]{\frac{\partial^2 #1}{\partial #2^2}}
\newcommand{\ud}{\,\mathrm{d}}

\newcommand{\be}{\begin{equation}}
\newcommand{\ee}{\end{equation}}
\newcommand{\ba}{\begin{array}}
\newcommand{\ea}{\end{array}}

\usepackage{color}
\definecolor{Green}{rgb}{0,0.7,0}

\def \k{{\bm{k}}}
\def \r{{\bm{r}}}

\begin{document}

\title{\color{black}Tunable orbital susceptibility in  $\alpha$-${\cal T}_3$ tight-binding models}
\author{F \textsc{Pi\'echon}$^{1}$, J-N Fuchs$^{1,2}$, A Raoux$^{1},^{3}$ and G Montambaux$^{1}$}
\address{$^1$ Laboratoire de Physique des Solides, CNRS UMR 8502, Universit\'e Paris-Sud, F-91405 Orsay Cedex, France \\
$^2$ Laboratoire de Physique Th\' eorique de la Mati\` ere Condens\' ee, CNRS UMR 7600, Univ. Pierre et Marie Curie 4, place Jussieu, 75252 Paris Cedex 05, France\\
$^3$ D\' epartement de Physique, \' Ecole Normale Sup\'erieure, 24 rue Lhomond, 75005 Paris, France}

\ead{piechon@lps.u-psud.fr}
\begin{abstract}
We study the importance of interband effects on the orbital susceptibility of three bands $\alpha$-${\cal T}_3$ tight-binding models.
The particularity of these models is that the coupling between the three energy bands (which is encoded in the wavefunctions properties)
can be tuned (by a parameter $\alpha$) without any modification of the energy spectrum.
Using the gauge-invariant perturbative formalism that we have recently developped~\cite{Raoux14b},
we obtain a generic formula of the orbital susceptibility of $\alpha$-${\cal T}_3$ tight-binding models.
Considering then three characteristic examples that exhibit either Dirac, semi-Dirac or quadratic band touching,
we show that by varying the parameter $\alpha$ and thus the wavefunctions interband couplings, it is possible to drive
a transition from a diamagnetic to a paramagnetic peak of the orbital susceptibility at the band touching.
In the presence of a gap separating the dispersive bands, we show that the susceptibility inside
the gap exhibits a similar dia to paramagnetic transition.
\end{abstract}

%
\section{Introduction}
\medskip

The orbital magnetic susceptibility of free electrons was computed long ago by Landau \cite{Landau30} and was found to be diamagnetic.
Subsequently, Peierls extended Landau's result to the case of a single band tight-binding model. He found a formula for the orbital susceptibility
that only depends on the zero-field band energy spectrum. This result already showed  that the band structure can have a strong
influence on the magnetic response of crystals \cite{Peierls33}. For example, near a saddle point of the dispersion relation, the Peierls orbital susceptibility becomes paramagnetic \cite{Vignale91}.
However, one important effect was left out, namely the coupling between the bands in the case of several bands. A striking example is the possibility of having
a finite orbital susceptibility inside the gap of a band insulator at zero temperature.
This was understood as an inter-band effect by Fukuyama and Kubo on the example of bismuth~\cite{Fukuyama70}. Fukuyama also provided a compact and quite general linear-response
formula that includes interband effects \cite{Fukuyama71}. However, despite its many successes, this formula does not work for non-separable tight-binding models.
The aim of this paper is to present orbital susceptibility results obtained from an exact linear response formula  that we recently derived  for tight-binding models \cite{Raoux14b}.
In order to show the importance of inter-band (or band coupling) effects, we  consider here a family of tunable three band tight-binding models constructed on a  $\mathcal T_3$ lattice (or \textit{dice} lattice)
and that we call $\alpha$-$\mathcal{T}_3$ \cite{Raoux14}. The dice model was first introduced \cite{Sutherland86} as a simple 2-dimensional model that exhibits localized and extended states as the same time.
It also presents some peculiar properties under strong magnetic field \cite{Vidal98}.
These $\alpha$-$\mathcal{T}_3$ models depend on a real parameter $\alpha$ and have the important property that their zero-field energy spectrum --which is essentially that of graphene with an additional flat band--
is independent of $\alpha$.
However, the parameter $\alpha$ has a strong influence on the zero-field eigenstates and therefore on the Berry curvature.
This influence is revealed by the orbital susceptibility that changes sign as a function of $\alpha$.

The paper is organized as follows. In \S 2 we present the three bands $\alpha$-${\cal T}_3$ tight-binding models and characterize the key properties of
their energy spectrum and wavefunctions.
In \S 3, using the gauge-invariant perturbative formalism that we have recently developped~\cite{Raoux14b},
we provide a generic formula for the orbital susceptibility of $\alpha$-${\cal T}_3$ tight-binding models.
In \S 4 we apply this susceptibity formula to three characteristic examples
of $\alpha$-${\cal T}_3$ tight-binding models that exhibit respectively Dirac, semi-Dirac and quadratic band touching at low-energy.
In \S 5 we summarize the main results of this study.

\section{Tight-binding models on the ${\cal T}_3$ lattice}
\medskip

\begin{figure}[t]
\begin{center}
{\includegraphics[width=14.cm]{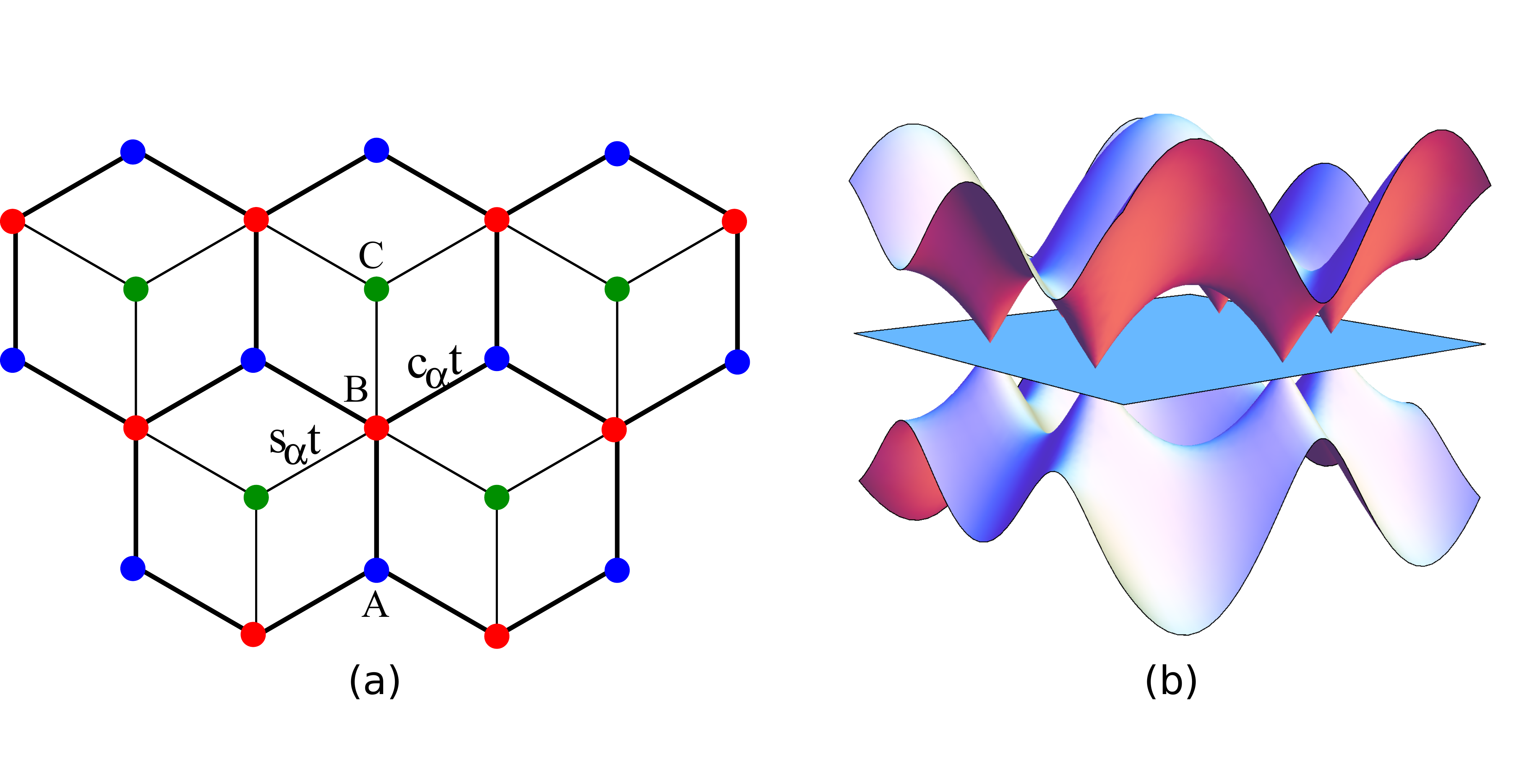}}
\caption{(Color online). $\alpha$-${\cal T}_3$ model: (a) The dice or ${\cal T}_3$ lattice is constituted by two interpenetrating honeycomb lattices.
Thick links, nearest-neighbor hoppings $c_{\alpha} t$ from sites $A$ to sites $B$ ($c_{\alpha}=\frac{1}{\sqrt{1+\alpha^2}}$).
Thin links, nearest-neighbor hoppings $s_{\alpha} t$ from sites $C$ to sites $B$ ($s_{\alpha}=\frac{\alpha}{\sqrt{1+\alpha^2}}$).
By varying $\alpha$, the model interpolates between honeycomb ($\alpha=0$) and dice ($\alpha=1$). Potential term $+\Delta$ on sites $A,C$ and $-\Delta$ on sites $B$.
(b) Energy bands dispersion in $\k$ space, for $\Delta=0$: two dispersive bands and one flat band. This spectrum does not depend on $\alpha$.}
\label{fig:T3lattice}
\end{center}
\end{figure}

Starting from the honeycomb lattice with two sites $(A,B)$ per unit cell, the ${\cal T}_3$ or dice lattice
is obtained by connecting additional ($C$) sites at the center of each hexagon to the $B$ sites (see Fig.~\ref{fig:T3lattice}).
The dice lattice is thus a triangular Bravais lattice with three sites  $(A,B,C)$ per unit cell.
We consider tight-binding models that consist of spinless electrons hopping on this lattice. In its simplest form,
we allow for a constant onsite potential term $+\Delta$ on sites $A,C$ and $-\Delta$ on sites $B$ and
an isotropic nearest-neighbors hopping with amplitude $c_{\alpha} t$ from $A$ to $B$ and $s_{\alpha} t$ from $C$ to $B$ with
$c_{\alpha}=\frac{1}{\sqrt{1+\alpha^2}}$, $s_{\alpha}=\frac{\alpha}{\sqrt{1+\alpha^2}}$ such that $c_\alpha ^2+s_\alpha ^2=1$.
The real space representation of the corresponding Hamiltonian follows as
\be
\label{rhamiltonian}
\ba{ll}
h=\sum_{\r_B}& \ [c_{\alpha}t (\delta_{\r_B-\bm \delta_1,\r_A}+\delta_{\r_B-\bm \delta_2,\r_A}+\delta_{\r_B-\bm \delta_3,\r_A})|\r_B\rangle\langle \r_A|\\
&+s_{\alpha}t (\delta_{\r_B+\bm \delta_1,\r_C}+\delta_{\r_B+\bm \delta_2,\r_C}+\delta_{\r_B+\bm \delta_3,\r_C})|\r_B\rangle\langle \r_C|] +\rm{h.c} ,
\ea
\ee
where $\bm \delta_1,\bm \delta_2,\bm \delta_3$ are the three vectors connecting nearest-neighbors sites.
We assume that the localized orbital basis is orthogonal ($\langle \r_{j'}|\r_j\rangle=\delta_{\r_{j'},\r_j}$) such that the position operator is purely diagonal
($\r=\sum_{\r_{j=A,B,C}} \r_j|\r_j\rangle\langle \r_j|$).
Introducing the Bloch states basis $|\k_{j=A,B,C}\rangle=\sum_{\r_{j}} e^{i\k \r_{j}}|\r_j \rangle$, for each wavevector $\k=(k_x,k_y)$,
the Bloch Hamiltonian matrix associated to this $\alpha$-${\cal T}_3$ tight-binding model reads
\be
h_\k =
\left(
\begin{array}{ccc}
            \Delta &  c_\alpha f_\k & 0 \\
            c_\alpha f_\k^*   & -\Delta &  s_\alpha  f_\k \\
            0 &    s_\alpha f_\k^*& \Delta \\
\end{array}
\right) \, ,
\label{hamiltonian}
\ee
where $f_\k=| f_\k| e^{-i\theta_\k}= t(e^{-i \k . \bm \delta_1}  + e^{-i \k . \bm \delta_2} + e^{-i \k . \bm \delta_3}) $.
In the following, we will also consider generalized versions of this model that essentially consist in a modified $f_\k$. This
 class of models interpolates between the honeycomb ($\alpha=0$) and the isotropic dice lattice ($\alpha=1$).

The remarkable and interesting property of this class of models is that the energy band spectrum does not depend on $\alpha$ (see Fig.~\ref{fig:T3lattice})
whereas the eigenfunctions do. More quantitatively, the energy spectrum consists of two dispersive bands $\epsilon_{\pm,\k}=\pm \sqrt{\Delta^2+|f_\k|^2}$
and a flat band at energy $\epsilon_{0}=\Delta$. For $\Delta>0$, the corresponding Bloch eigenfunctions $|s_{\alpha,\k}\rangle$  ($s=\pm,0$) read
\be
|+_{\alpha,\k}\rangle=\frac{\sqrt{1+c_{\k}}}{\sqrt{2}}
\left(\ba{c}
c_{\alpha}  e^{-i \theta_\k}\\
\\
\frac{s_{\k}}{1+c_{\k}}\\
\\
s_{\alpha}  e^{i \theta_\k}
\ea \right); \
|-_{\alpha,\k}\rangle=\frac{s_{\k}}{\sqrt{2(1+c_{\k})}}
\left(\ba{c}
c_{\alpha}  e^{-i \theta_\k}\\
\\
-\frac{1+c_{\k}}{s_{\k}}\\
\\
s_{\alpha}  e^{i \theta_\k}
\ea \right); \
|0_{\alpha,\k}\rangle=
\left(\ba{c}
s_{\alpha}  e^{-i \theta_\k}\\
\\
0\\
\\
 -c_{\alpha}  e^{i \theta_\k}
\ea \right)
\ee
where $c_\k=\frac{\Delta}{\sqrt{\Delta^2+|f_\k|^2}}$ and $s_{\k}=\frac{|f_\k|}{\sqrt{\Delta^2+|f_\k|^2}}$ with $c_\k ^2+s_\k ^2=1$.
The Berry connection ${\mathcal  A}_{\k}^s=\langle s_{\alpha,\k}|i\nabla_{\k} |s_{\alpha,\k}\rangle$ of each band is then readily obtained as
\be
{\mathcal A}_{\k}^0=-\frac{1-\alpha^2}{1+\alpha^2}\nabla_\k \theta; \ \ \
{\mathcal A}_{\k}^+=-\frac{1+c_\k}{2}{\mathcal A}_{\k}^0; \ \ \
{\mathcal A}_{\k}^-=-\frac{s_\k ^2}{2(1+c_\k)}{\mathcal A}_{\k}^0,
\label{berry}
\ee
such that quite generally ${\mathcal A}_{\alpha,\k}^0+{\mathcal A}_{\alpha,\k}^++{\mathcal A}_{\alpha,\k}^-=0$.
From this Berry connection it is possible to obtain, for each band, the so-called Berry curvature $\Omega_\k^s=\nabla_\k \times {\mathcal  A}_{\k}^s$.
This Berry curvature usually presents some strongly peaked structure near specific $\k^*$ points where interband coupling is strong, even
if there is a gap separating the bands. Moreover,   by considering a k-space closed orbit ${\mathscr C}_{\k^*}$ around such a $\k^*$ point,
it is possible to calculate a Berry phase that measures the winding of the phase $\theta_\k$ around such points $\k^*$.
More precisely, considering closed orbits ${\mathscr C}_{\k^*}(\varepsilon)$ that corresponds to a constant energy $\varepsilon=\sqrt{\Delta^2+|f_\k|^2}$,
we deduce that for each band the Berry phase $\Phi_{\k^*}^s= \oint_{{\mathscr C}_{\k^*}} {\rm{d}} \k  \cdot {\mathcal  A}_{\k}^s$ accumulated along such an orbit is given respectively by
$\Phi_{\k^*}^0=-  \Phi_B   W_{{\mathscr C}_{\k^*}}$  with 
$\Phi_{\k^*}^{\pm}=-\frac{1}{2}(1\pm \frac{\Delta}{\epsilon})\Phi_{\k^*}^0$ where $\Phi_B =\pi \frac{1-\alpha^2}{1+\alpha^2}$ and 
$W_{{\mathscr C}_{\k^*}}= \oint_{{\mathscr C}_{\k^*}} {\rm{d}} k\cdot\nabla_\k \theta/2\pi$
is the winding number of the orbit such that $W_{\mathscr C}=0$ for a closed orbit around a regular $\k^*$-point and $W_{{\mathscr C}_{\k^*}}=\pm 1$ for an orbit encircling a Dirac point $\k^*$~\cite{Raoux14,Fuchs10}.

From this perspective, the main interest of the $\alpha$-${\cal T}_3$ models described by Eq.(\ref{hamiltonian}) is that all these Berry quantities (connections, curvature and phase) are proportional
to $\Phi_B= \pi \frac{1-\alpha^2}{1+\alpha^2}$ (see Eq.(\ref{berry})). As a consequence, the interband effects that are encoded by these Berry quantities are reduced upon increasing the value of $\alpha$ from $0$
and totally vanish for $\alpha=1$. Anticipating on what follows, we point out that our results for the orbital susceptibility of $\alpha$-${\cal T}_3$ models provide clear evidence that all
interband effects are not only encoded by the Berry quantities.


\section{Orbital susceptibility formula}
\medskip

For time reversal systems which we consider in the following, the orbital magnetization vanishes and the orbital susceptibility  $\chi_\mathrm{orb}(\mu,T)$
is the first measure of the sensitivity of the energy spectrum when it is placed in a uniform static perpendicular magnetic field ${\bm B}=B {\bm u}_z$:
\begin{equation}
\label{eq:def_chi}
\chi_\mathrm{orb}(\mu,T)=-\frac{\mu_0}{S}\left.\ddpar{\Omega}{B}\right|_{B=0}
\end{equation}
where $S$ is the area of the system, $T$ is the temperature, $\mu$ is the chemical potential, $\mu_0=4\pi.10^{-7}$S.I. and $\Omega(T,\mu,B)$ is the grand canonical potential of the non-interacting electrons gas
\begin{equation}
\label{eq:def_GP}
\Omega(T,\mu,B)=-T\int_{-\infty}^{+\infty} \ln\left(1+e^{-(\omega-\mu)/T}\right)\rho(\omega,B)\,\mathrm{d}\omega,
\end{equation}
with $\rho(\omega,B)$ the field dependent density of states (Dos)
\begin{equation}
\label{eq:def_g}
\rho(\omega,B)=-\frac{1}{\pi}\Im m \Tr\,G(\omega+{\rm{i}}0^+,B)
\end{equation}
where $G(\omega,B)$ is the field dependent retarded Green's function (we use $k_\mathrm{B}=1$ and $\hbar=1$).

In order to obtain the susceptibility, it is thus necessary to calculate the density of states $\rho(\omega,B)$ or the Green's function $G(\omega,B)=(\omega {\rm{I}}-h(B))^{-1}$ associated
to the field dependent Hamiltonian $h(B)$.  For the tight-binding model (\ref{rhamiltonian}), $h(B)$ is obtained by multiplying the real space hopping amplitudes by a gauge and position dependent Peierls phase:
\be
\label{rhamiltonianB}
\ba{ll}
h(B)=\sum_{\r_B}& \ [c_{\alpha}t (\delta_{\r_B+\bm \delta_1,\r_A}+\delta_{\r_B+\bm \delta_2,\r_A}+\delta_{\r_B+\bm \delta_3,\r_A})e^{-i\varphi_{\r_A,\r_B}}|\r_B\rangle\langle \r_A|\\
&+s_{\alpha}t (\delta_{\r_B-\bm \delta_1,\r_C}+\delta_{\r_B-\bm \delta_2,\r_C}+\delta_{\r_B-\bm \delta_3,\r_C})e^{-i\varphi_{\r_C,\r_B}}|\r_B\rangle\langle \r_C|] +\rm{h.c} ,
\ea
\ee
where $\varphi_{\r,\r'}=\frac{e}{\hbar}\int_{\r} ^{\r'} {\bm A}\cdot \ud{\bm l}$ with ${\bm A}(\r)$ the gauge dependent vector potential associated to the uniform magnetic field.
Starting from the Hamiltonian $h(B)$ there are essentially two approaches to calculate $\rho(\omega,B)$.

\begin{figure}[t]
\begin{center}
{\includegraphics[width=7.cm]{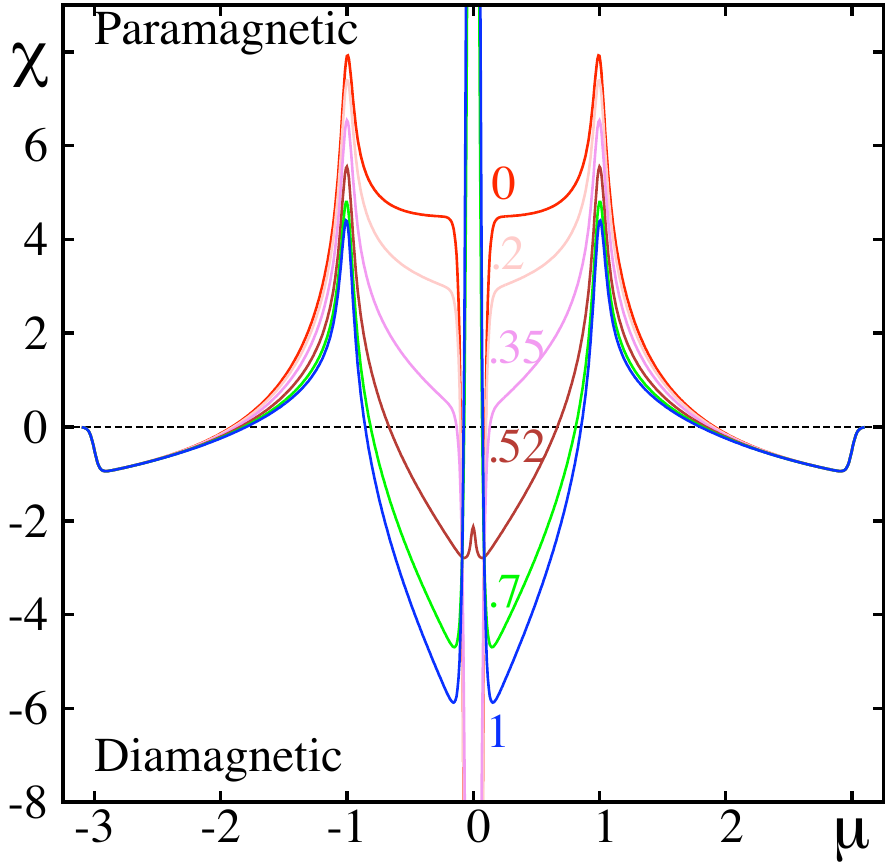}}
\caption{(Color online). Orbital susceptibility $\chi(\mu,T)$ obtained from the numerically computed Hofstadter spectrum~\cite{Raoux14}.
$\chi$ (in units of the Landau band edge value $|\chi_L|=(\mu_0/16\pi)(e^2t a^2/\hbar)$) as a function of the chemical potential $\mu$
(in units of $t$) in the whole band for various $\alpha$ as indicated and for
a temperature $T=0.02t$. At $\mu=0$ there is a transition from a diamagnetic peak for $\alpha=0$ (red: graphene)
to a paramagnetic peak for $\alpha=1$ (blue: dice). Because of a sum rule,
the orbital response at zero doping is systematically compensated
by an opposite response at finite doping.}
\label{fig:hofstadter}
\end{center}
\end{figure}

The first approach consists in computing the exact magnetic field dependent energy spectrum $\epsilon_n(B)$ associated to $h(B)$; this leads to the so-called Hofstadter butterfly spectrum.
Using such an approach, we have recently studied the orbital susceptibility of the $\alpha$-${\cal T}_3$ model (\ref{hamiltonian}) for the case $\Delta=0$~\cite{Raoux14}.
In particular we have shown that the orbital susceptibility $\chi_\mathrm{orb}(\mu,T)$ strongly varies with the parameter $\alpha$. More precisely, at the Dirac point, $\chi_\mathrm{orb}(\mu=0,T)$
exhibits a continuous transition from a diamagnetic peak for $\alpha=0$ (honeycomb-graphene)
to a paramagnetic peak for $\alpha=1$ (dice) (see Fig.~\ref{fig:hofstadter}). Away from the Dirac point, an opposite transition from paramagnetism to diamagnetism takes place such
that a sum rule $\int \ud \mu \chi_\mathrm{orb}(\mu,T)=0$ is preserved for all $\alpha$~\cite{Raoux14}. Despite these interesting results, such an approach suffers
from being essentially numerical and thus it does not allow to fully understand how the dependence on the parameter $\alpha$ enters in the susceptibility.

In order to better understand to role of the parameter $\alpha$, we have developped a second approach which consists in calculating the susceptibility $\chi_\mathrm{orb}(\mu,T)$
using our recently established gauge-invariant perturbative response formula~\cite{Raoux14b}:
\begin{equation}
\chi_\mathrm{orb}=-\frac{\mu_0 e^2}{12\hbar^2}\frac{\Im m}{\pi S}\int_{-\infty}^{+\infty}  \! \! \ud \omega \
n_\mathrm{F}(\omega) \  \Tr\left\{gh^{xx}gh^{yy}-gh^{xy}gh^{xy}- 4(gh^xgh^xgh^ygh^y-gh^xgh^ygh^xgh^y)\right\},
\label{eq:chi1}
\end{equation}
where $n_\mathrm{F}(\omega)=1/(e^{\frac{\omega-\mu}{T}}+1)$ is the Fermi function, $g(\omega)=(\omega {\rm{I}}-h)^{-1}$ is the retarded
Green's function associated to the zero-field Hamiltonian and $h^x=[x,h]$, $h^{xx}=[x,[x,h]]$ are the single and double commutators of the position operator with the zero-field Hamiltonian.
As discussed in~\cite{Raoux14b}, the expression (\ref{eq:chi1}) is equivalent to a recent formula derived for graphene~\cite{Gomez-Santos11}
but it differs from the well-known Fukuyama formula~\cite{Fukuyama71}. We stress however that when $h^{xy} \ne 0$, only formula (\ref{eq:chi1}) fully agrees with the susceptibility
results obtained from direct numerical computation of the corresponding Hofstadter butterfly spectrum. Moreover, we remind that formula (\ref{eq:chi1}) is valid not only for Bloch electrons
in infinite crystals but it also applies to disordered and finite systems as well.

In the present paper, we restrict to multiband Bloch electrons in infinite crystals. In that situation, the trace operator
is rewritten $\Tr(\bullet)=\sum_{\bm k}\mathrm{tr}(\bullet)=S\int \frac{\ud^2k}{4\pi^2}\mathrm{tr}(\bullet)$ where the integration
is performed over the first Brillouin zone (BZ) and $\mathrm{tr}(\bullet)$ is the partial trace operator on the band index. Accordingly, the susceptibility formula follows as
\begin{equation}
\chi_\mathrm{orb}=-\frac{\mu_0 e^2}{12}\frac{\Im m}{\pi }\int_{-\infty}^{+\infty} \ud \omega \int \frac{\ud^2k}{4\pi^2} \
n_\mathrm{F}(\omega)  [ U_\k(\omega)-4V_\k(\omega)],
\label{eq:chi2}
\end{equation}
where
\begin{equation}
\ba{l}
U_\k(\omega)=\mathrm{tr}\left\{(gh^{xx}gh^{yy}-gh^{xy}gh^{xy})_\k\right\},\\
V_\k(\omega)=\mathrm{tr}\left\{(gh^xgh^xgh^ygh^y-gh^xgh^ygh^xgh^y)_\k\right\},
\ea
\label{eq:uvk}
\end{equation}
with $g_\k(\omega)=(\omega{\rm{I}}-h_\k)^{-1}$ the Green's function matrix associated to the zero-field Bloch Hamiltonian matrix and
$h_\k ^j=\frac{\partial h_\k}{\partial {k_j}}$, $h_\k ^{ij}=\frac{\partial^2 h_\k}{\partial {k_i}\partial{k_j}}$ with $(i,j)\in (x,y)$.
As detailed in the appendix, for the class of models described by Eq.(\ref{hamiltonian}),
by defining the two components vectors ($\vec{x}\equiv(x_1,x_2)$)
\begin{equation}
\ba{lll}
\vec{f}_\k =(\Re e f_\k, \Im m f_\k),&
\vec{f}_\k ^i= (\frac{\partial \Re e f_\k}{\partial {k_i}},\frac{\partial \Im m f_\k}{\partial {k_i}}),&
\vec{f}_\k ^{ij}=(\frac{\partial^2\Re e f_\k}{\partial {k_i}\partial{k_j}},\frac{\partial^2 \Im m f_\k}{\partial {k_i}\partial{k_j}}),
\ea
\label{veca}
\end{equation}
and the quantities
\begin{equation}
\ba{llll}
u_\k^j=\frac{\vec{f}_\k^j \cdot \vec{f}_\k}{|f_\k|},& v_\k^j=\frac{(\vec{f}_\k^j\times \vec{f}_\k)}{|f_\k|}, &
u_\k^{ij}=\frac{\vec{f}_\k^{ij} \cdot \vec{f}_\k}{|f_\k|}, & v_\k^{ij}=\frac{(\vec{f}_\k^{ij} \times \vec{f}_\k)}{|f_\k|},
\ea
\label{vecb}
\end{equation}
it is possible to obtain the following compact expressions for $U_\k(\omega)$ and $V_\k(\omega)$:
\begin{equation}
\label{eq:uvkb}
\ba{l}
U_\k(\omega)=2g_+g_-[(u_\k^{xx}u_\k^{yy}-u_\k^{xy}u_\k^{xy})+(v_\k^{xx}v_\k^{yy}-v_\k^{xy}v_\k^{xy})]
+4g_+^2g_-^2(u_\k^{xx}u_\k^{yy}-u_\k^{xy}u_\k^{xy}) |f_\k|^2,\\
V_\k(\omega)=(u_\k^xv_\k^y-u_\k^yv_\k^x)^2\left[g_+^2g_-^2[1+3\left(\frac{1-\alpha^2}{1+\alpha^2}\right)^2]
+4g_+^3g_-^3|f_\k|^2\right].
\ea
\end{equation}
where $g_{\pm}(\omega)=1/(\omega-\epsilon_{\pm,\k})$.
Interestingly, these last expressions show that $U_\k(\omega)$ and $V_\k(\omega)$ have poles only on the two dispersive bands as if the flat band did not play any role.
In fact, the implicit effect of the flat band is essentially encoded in the $\alpha$ dependent term appearing in $V_\k(\omega)$. However it is quite remarkable that $U_\k(\omega)$ is totally independent of $\alpha$.

\section{Orbital susceptibility of low-energy  $\alpha$-${\cal T}_3$ models}
\medskip

\begin{figure}[t]
\begin{center}
\includegraphics[width=7cm]{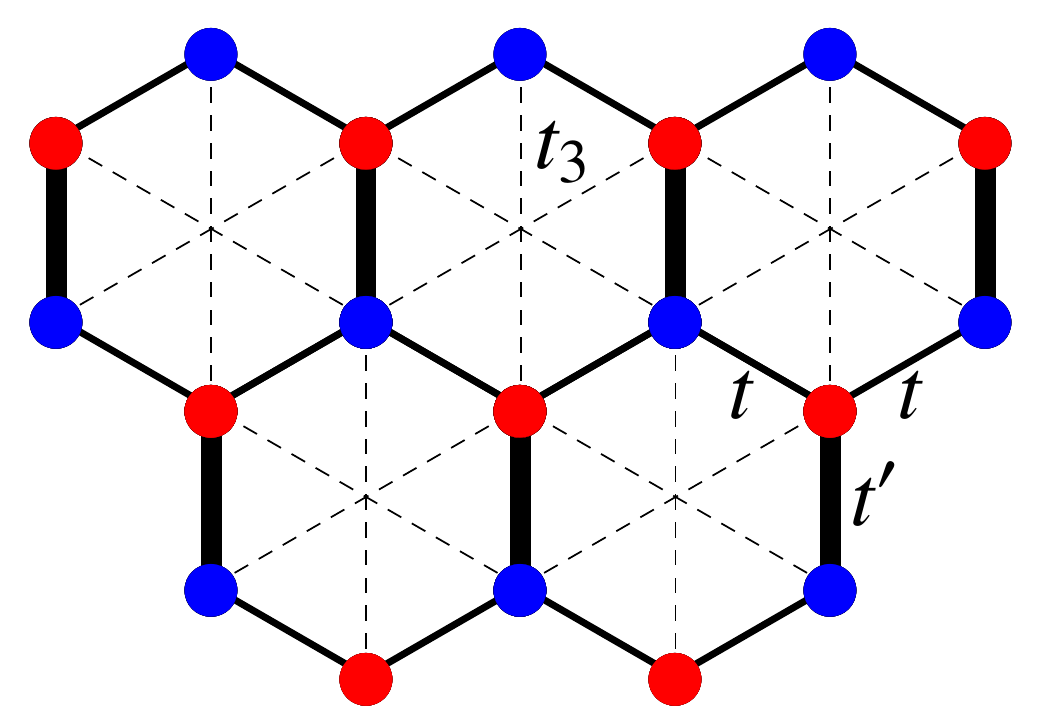}
\caption{(color online) Honeycomb lattice model that determines the form $f_\k$:  $t$ (full lines) and $t_3$ (dashed lines) respectively first and third nearest-neighbor hopping amplitudes.
The two sites in the unit cell are called $A$ and $B$ and are shown as blue and red dots. In a uniaxially compressed honeycomb lattice,
there are two different values for the nearest-neighbor hopping amplitudes: $t$ for thin (non--vertical) lines and $t'\geq t$ for thick (vertical) lines.}
\label{fig:honeycomb}
\end{center}
\end{figure}

In this part of the paper we use equations (\ref{vecb},\ref{eq:uvkb}) to calculate the susceptibility of $\alpha$-${\cal T}_3$ model for three different $f_\k$.
In order to allow for a full analytical calculation, for each example we consider the low-energy effective model Hamiltonian that correctly describes the energy spectrum
near the minimum of the corresponding $|f_\k|$. The three distinct forms of $f_\k$ that we consider are defined on the honeycomb lattice model illustrated on Fig.~\ref{fig:honeycomb},
that corresponds to the $\alpha=0$ limit in which the $C$ sites are completely decoupled from $A,B$ sites. The $C$ sites are not shown on Fig.~\ref{fig:honeycomb} for simplicity.

\subsection{$\alpha$-${\cal T}_3$ graphene model}
\medskip

We first consider the usual isotropic model of graphene which corresponds to $t'=t$ and $t_3=0$ in Fig.~\ref{fig:honeycomb}. In this situation,
\begin{equation}
 f_\k= t(e^{-i \k . \bm \delta_1}  + e^{-i \k . \bm \delta_2} + e^{-i \k . \bm \delta_3})=t(
 e^{-ik_ya}+2e^{ik_ya/2}\cos(\sqrt3k_xa/2)) \\
\end{equation}
For $\Delta=0$, the corresponding $\alpha$-${\cal T}_3$ energy spectrum (see Fig.~\ref{fig:T3lattice}(b)) exhibits linear band touching at the two inequivalent corners $\pm {\bm K}$ of the Brillouin zone.
Introducing the valley index $\xi=\pm$, the effective low-energy model is obtained by expanding $f_{\bm k}$ near $\xi {\bm K}$: $f_{\xi {\bm K} +\k}\simeq v(\xi k_x-ik_y)$ with the velocity $v=\frac{3ta}{2}$
where $a$ is the nearest-neighbor distance.
Hereafter to simplify the notations of most equations we define the pseudo wavevector $\kappa_{x,y}=v k_{x,y}$.
For each valley, the low-energy $\alpha$-${\cal T}_3$ Hamiltonian reads:
\be
\label{lowhamil1}
h_{\k} =
\left(
\begin{array}{ccc}
            \Delta&  c_\alpha (\xi \kappa_x-i\kappa_y) & 0 \\
            c_\alpha (\xi \kappa_x+i\kappa_y) & -\Delta&  s_\alpha   (\xi \kappa_x-i\kappa_y)\\
            0 &    s_\alpha (\xi \kappa_x+i\kappa_y)& \Delta \\
\end{array}
\right) \, .
\ee
It is then immediate to obtain
\be
\ba{llll}
\vec{f}=(\xi \kappa_x, -\kappa_y),&\vec{f}^x=(\xi,0),& \vec{f}^y=(0,-1),&\vec{f}^{ij}=0.
\ea
\ee
and
\be
\ba{lll}
u^x= \xi v_y=\frac{\kappa_x}{\sqrt{\kappa_x^2+\kappa_y^2}}, & u^y=-\xi v_x=-\frac{\kappa_y}{\sqrt{\kappa_x^2+\kappa_y^2}},&
u^{ij}=v^{ij}=0,
\ea
\ee
from which we deduce
\begin{equation}
\label{eq:uvkgraphene}
\ba{l}
U_\k(\omega)=0,\\
V_\k(\omega)= \left[\frac{1}{(\omega^2-(\Delta^2+\kappa_x^2+\kappa_y^2))^2}[1+3\left(\frac{1-\alpha^2}{1+\alpha^2}\right)^2]
+\frac{4(\kappa_x^2+\kappa_y^2)}{(\omega^2-(\Delta^2+\kappa_x^2+\kappa_y^2))^3}\right]
\ea
\end{equation}
Substituting these expressions in (\ref{eq:chi2}), summing over the two valleys
and noting  $\kappa_x=\varepsilon \cos{\theta}$, $\kappa_y=\varepsilon \sin{\theta}$, we obtain
\begin{equation}
\label{eq:chigraphene}
\ba{ll}
\chi_\mathrm{orb}(\mu)&=\frac{2\mu_0 e^2}{12}\left(\frac{4}{4\pi^2}\right)\frac{\Im m}{\pi }\int_{-\infty}^{+\infty} \ud \omega \ n_\mathrm{F}(\omega)
\int_0 ^{2\pi} \ud \theta \\
& \times \int_0 ^{\infty} \ud \varepsilon \ \varepsilon
\left[\frac{1}{(\omega^2-(\Delta^2+\varepsilon^2))^2}[1+3\left(\frac{1-\alpha^2}{1+\alpha^2}\right)^2] +\frac{4\varepsilon^2}{(\omega^2-(\Delta^2+\varepsilon^2))^3}\right].
\ea
\end{equation}
Integrating over variables and following the procedure described in appendix B, we finally obtain
\begin{equation}
\label{eq:chigrapheneb}
\chi_\mathrm{orb}(\mu=0)=\frac{\mu_0 e^2 v^2}{2\pi}\left[\left(\frac{1-\alpha^2}{1+\alpha^2}\right)^2-\frac{1}{3}\right]
\ba{l}
\frac{n_F(\Delta)-n_F(-\Delta)}{2\Delta}.
\ea
\end{equation}
which gives the correct behaviour for $\Delta=0$.
For $\alpha=0$, Eq.(\ref{eq:chigrapheneb}) with $\Delta=0$ recovers the famous diamagnetic peak originally found by McClure~\cite{McClure56},
whereas Eq.(\ref{eq:chigrapheneb}) with $\Delta\not=0$ reproduces the in-gap diamagnetic \textit{\textit{plateau}} first obtained by Koshino and Ando~\cite{Koshino07}.
Note however that these authors obtained  their results by performing a low-field expansion of the grand potential calculated from direct summation over
the effective Landau levels spectrum $\epsilon_n(B)= \pm \epsilon_B \sqrt{|n|}$~\cite{McClure56} or $\epsilon_n(B)= \pm \sqrt{\Delta^2+ \epsilon_B^2 {|n|}}$~\cite{Koshino07} where $\epsilon_B=v \sqrt{2e B}$.
For $\alpha \ne 0$, Eq.(\ref{eq:chigrapheneb}) is also coherent with the results of such methods,  albeit now with
$\alpha$-dependent effective Landau levels spectrum $\epsilon_n(B)= \pm \epsilon_B \sqrt{|n+\gamma|}$ or $\epsilon_n(B)= \pm \sqrt{\Delta^2+\epsilon_B^2 |n+\gamma|}$ where  $n \in  \mathbb{Z}$  and $\gamma=\frac{\alpha^2}{1+\alpha^2}$~\cite{Raoux14}.
In this picture, using the relation $\gamma=\frac{1}{2}-\Phi_B/2\pi$ where $\Phi_B$
is the $\alpha$-dependent Berry phase of a Dirac cone~\cite{Raoux14,Fuchs10}, the variation of the susceptibility with $\alpha$ is interpreted as a consequence of the variation of the corresponding Berry phase.
As we already pointed out, for $\alpha=1$ the Berry phase vanishes and so do interband effects encoded by the Berry curvature.
Nevertheless, for $\alpha=1$ there is still a paramagnetic susceptibility \textit{plateau} inside the gap which is an indication that interband effects are still present.
This result implies that some important interband effects are not encoded by the Berry curvature.
More quantitatively we note that the $\alpha$ dependent prefactor of the susceptibility (\ref{eq:chigrapheneb}) changes sign at
$\alpha_c=\frac{\sqrt{3}-1}{\sqrt{2}}\simeq 0.518$. This signals
a transition from a diamagnetic peak/\textit{plateau} for $\alpha<\alpha_c$ to a paramagnetic peak/\textit{plateau} for $\alpha>\alpha_c$.
This value $\alpha_c= 0.518$ coincides with the numerical results obtained from computing the susceptibility with the Hofstadter spectrum~\cite{Raoux14}.
As a final remark, we note that for the low-energy Hamiltonian (\ref{lowhamil1}) which is linear in $\k$ and thus separable (i.e. $h^{xy}_\k=0$),
the Fukuyama formula~\cite{Fukuyama71} would have given the same results.

\subsection{$\alpha$-${\cal T}_3$ semi-Dirac model}
\medskip

\begin{figure}
\begin{center}
\begin{tabular}{cc}

{\includegraphics[width=4.5cm]{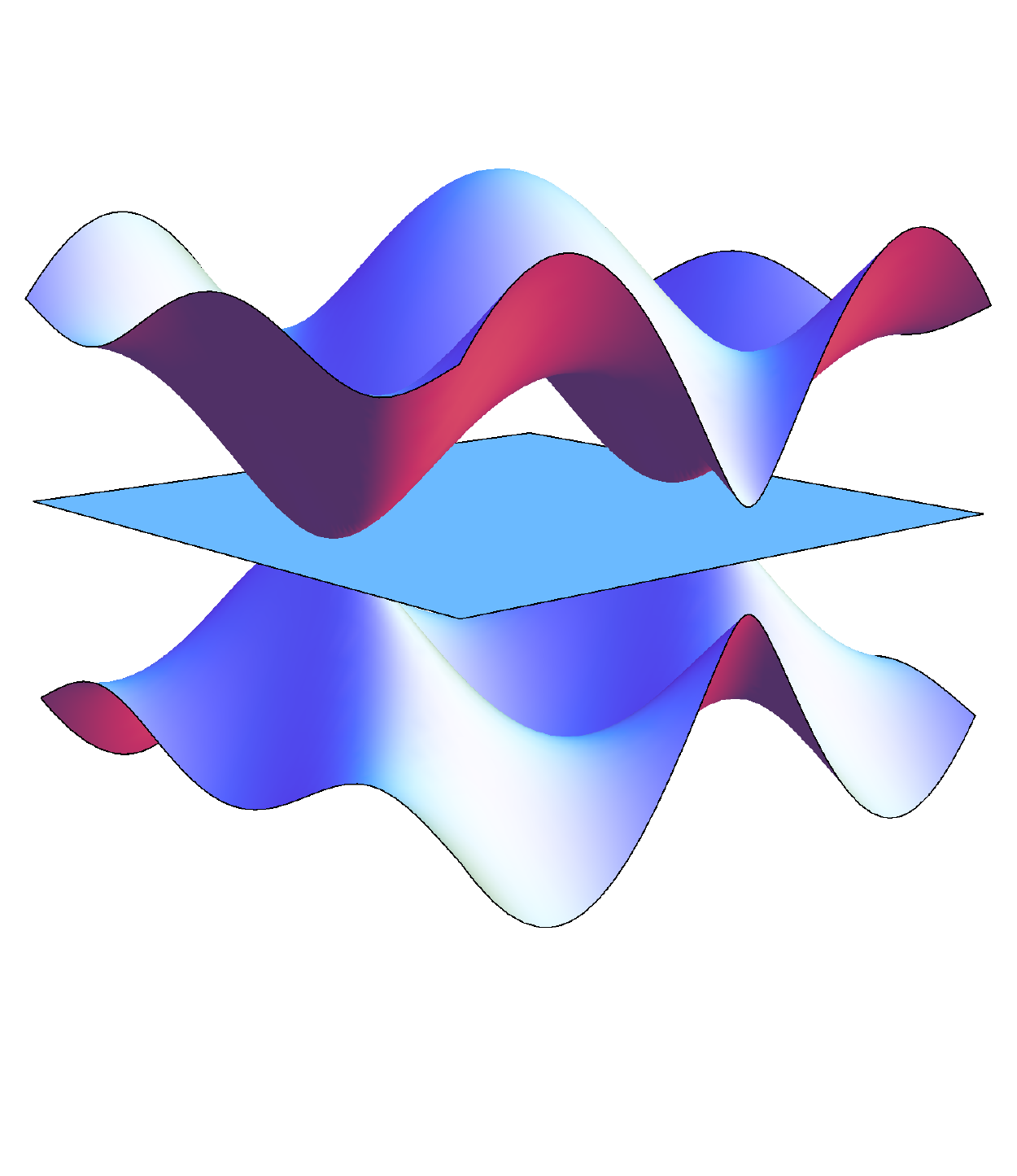}}&
{\includegraphics[width=4.cm]{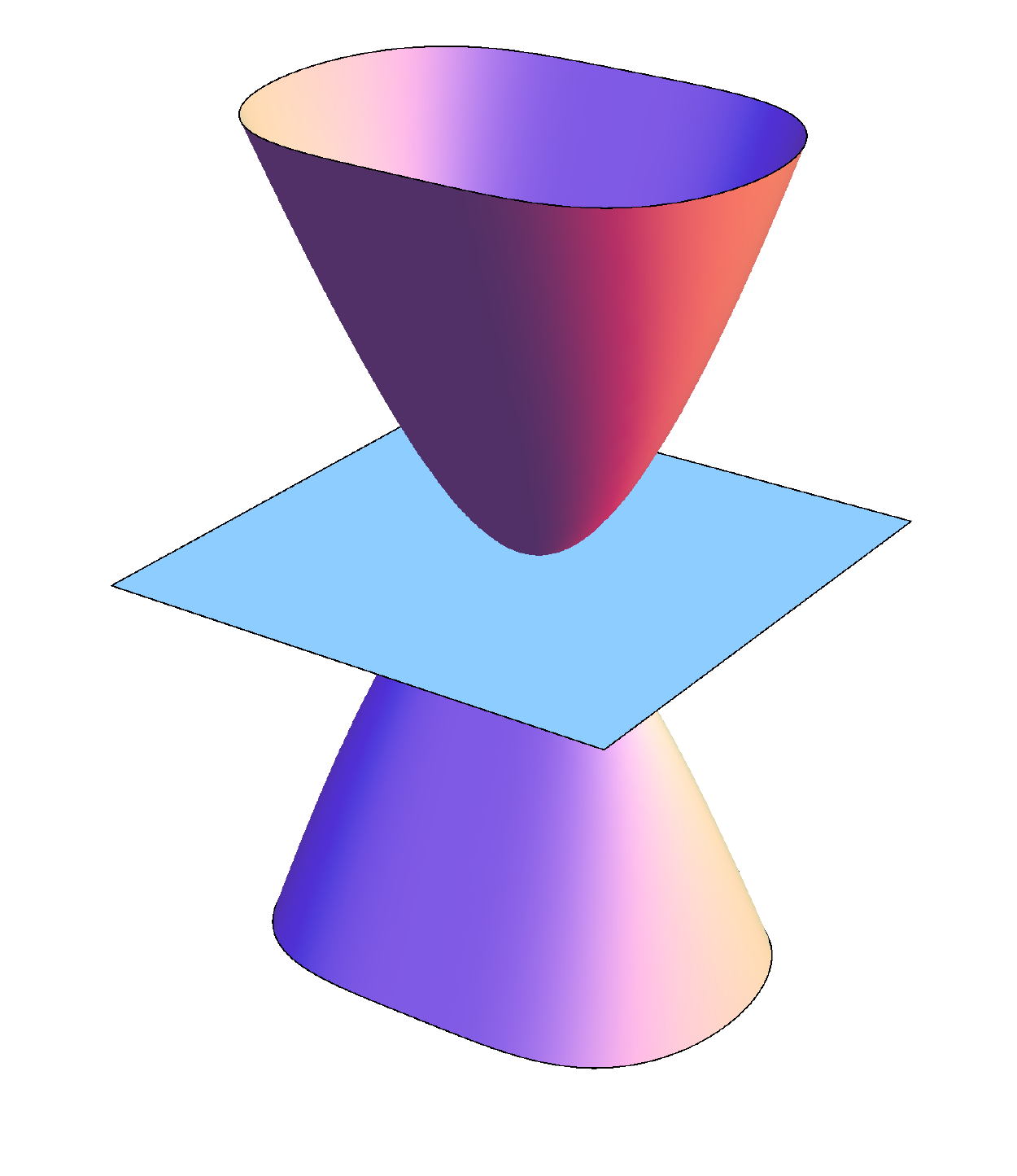}}\\
 \Large $(a)$&\Large$(b)$
 \end{tabular}
\caption{(Color online). $\alpha$-${\cal T}_3$ semi-Dirac model
energy bands dispersion in $\k$ space, for $\Delta=0$. (a) Full spectrum with semi-Dirac band touching of the two dispersive bands. (b) low-energy spectrum.}
\label{fig:t3merging.pdf}
\end{center}
\end{figure}

We now consider a model with $t'=2t$ and $t_3=0$ (see Fig. \ref{fig:honeycomb}). In that situation
\begin{equation}
 f_\k= t(2e^{-ik_ya}+2e^{ik_ya/2}\cos(\sqrt3k_xa/2))\\
\end{equation}
For $\Delta=0$, the corresponding $\alpha$-${\cal T}_3$ energy spectrum (see Fig.~\ref{fig:t3merging.pdf}) exhibits a semi-Dirac band touching
at one of the $M$ points of the Brillouin zone~\cite{Dietl08}. The low-energy model obtained by expanding near this point reads $f_{M+\k}\simeq \frac{k_x^2}{2m_*}-ivk_y$
with the effective mass $m_*=\frac{1}{ta^2}$ velocity $v=\frac{ta}{\hbar}$ and it features a linear-quadratic spectrum $\varepsilon_\k=\pm \sqrt{ (\frac{k_x^2}{2m_*})^2+(vk_y)^2}$.
As in previous example, to simplify the notations of most equations we define the pseudo wavevectors $\kappa_{x}=\frac{k_{x}}{\sqrt{2m_*}} $ and $\kappa_y=vk_y$.
The low-energy $\alpha$-${\cal T}_3$ Hamiltonian is:
\be
\label{lowhamil2}
h_{\k} =
\left(
\begin{array}{ccc}
            \Delta&  c_\alpha (\kappa_x^2-i\kappa_y) & 0 \\
            c_\alpha (\kappa_x^2+i\kappa_y) & -\Delta&  s_\alpha   (\kappa_x^2-i\kappa_y)\\
            0 &    s_\alpha (\kappa_x^2+i\kappa_y)& \Delta \\
\end{array}
\right) \, ,
\ee
such that
\be
\ba{lllll}
\vec{f}=(\kappa_x^2, -\kappa_y),& \vec{f}^x=2(\kappa_x,0),& \vec{f}^y=(0,-1)
\ea
\ee
and
\be
\ba{lllll}
u^x=\frac{2\kappa_x^3}{\sqrt{\kappa_x^4+\kappa_y^2}},& u^y=\frac{k_y}{\sqrt{\kappa_x^4+\kappa_y^2}},&
v^x=\frac{-2\kappa_x\kappa_y}{\sqrt{\kappa_x^4+\kappa_y^2}},& v^y=\frac{k_x^2}{\sqrt{\kappa_x^4+\kappa_y^2}},
\ea
\ee
from which we deduce
\begin{equation}
\label{eq:uvkmerging}
\ba{l}
U_\k(\omega)=0,\\
V_\k(\omega)= 4 \kappa_x ^2 \left[\frac{1}{(\omega^2-(\Delta^2+\kappa_x^4+\kappa_y^2))^2}[1+3\left(\frac{1-\alpha^2}{1+\alpha^2}\right)^2]
+\frac{4(\kappa_x^4+\kappa_y^2)}{(\omega^2-(\Delta^2+\kappa_x^4+\kappa_y^2))^3}\right].
\ea
\end{equation}

Writing now $\kappa_x^2=\varepsilon \cos \theta$ and $\kappa_y=\varepsilon \sin \theta$ with $\theta \in [0,\pi/2]$ we deduce
\begin{equation}
\label{eq:chimerging}
\ba{ll}
\chi_\mathrm{orb}(\mu)&=\frac{\mu_0 e^2}{12}\left(\frac{64}{4\pi^2}\right)\frac{\Im m}{\pi }\int_{-\infty}^{+\infty} \ud \omega \ n_\mathrm{F}(\omega)
\int_0^{\pi/2} \ud \theta \sqrt{\cos \theta} \\
&\times \int_0 ^{\infty} \ud \varepsilon  \ \varepsilon^{3/2} \
 \left[\frac{1}{(\omega^2-(\Delta^2+\varepsilon^2))^2}\left[1+3\left(\frac{1-\alpha^2}{1+\alpha^2}\right)^2\right]
+\frac{4\varepsilon^2}{(\omega^2-(\Delta^2+\varepsilon^2))^3}\right]\\
\ea
\end{equation}

Quite interestingly, apart from some prefactor, the main noticable change between equations (\ref{eq:chigraphene}) and  (\ref{eq:chimerging})
is the effective {\em density of states} that appears in the integral: for the $\alpha$-${\cal T}_3$ semi-Dirac model it is $\propto \varepsilon^{3/2}$
whereas it is $\propto \varepsilon$ in the $\alpha$-${\cal T}_3$ graphene model.
From formulae (\ref{appchi3},\ref{appchi5}) given in appendix B, we finally obtain:
\begin{equation}
\label{eq:chimergingb}
\chi_\mathrm{orb}(\mu)=-\frac{\mu_0 e^2 v}{\sqrt{2m_*}}\frac{\Gamma(\frac{3}{4})^2}{\pi^2 \sqrt{2\pi}}\left[\left(\frac{1-\alpha^2}{1+\alpha^2}\right)^2-\frac{1}{2}\right]
\times \left\lbrace
\ba{ll}
 \frac{1}{\sqrt{|\mu|}} & \Delta=0, T=0\\
&\\
\left(\frac{4\Gamma(\frac{5}{4})^2}{\sqrt{\pi}}\right)\frac{1}{\sqrt{\Delta}}& |\mu|<\Delta, T=0.
\ea
\right.
\end{equation}

For $\alpha=0$ and $\Delta=0$, the first expression coincides with the result  recently obtained in~\cite{Raoux14b} for a two band model at the semi-Dirac point.
For any $\alpha$ it predicts a diamagnetic peak that scales like $1/\sqrt{{\rm{min}}(\mu,T)}$ in the gapless case $\Delta=0$. In the presence of a gap, the second expression
shows that the susceptibility \textit{plateau} in the gap scales as $1/\sqrt{\Delta}$. As in the previous example, Eq.(\ref{eq:chimergingb}) predicts a diamagnetic to paramagnetic
transition for $\alpha\ge \alpha_c$ at a critical value $\alpha_c=\sqrt{2}-1\simeq 0.414$ which is smaller than in then previous example.
We have verified that this value $\alpha_c= 0.414$ agrees with the numerical results obtained from computing the susceptibility with the Hofstadter spectrum.
We stress that for the $\alpha$-${\cal T}_3$ semi-Dirac model, as yet, there is no exact analytical expression for the Landau levels $\epsilon_n(B)$ of the low-energy model. The   Landau levels  are related to the eigenvalues of a modified quartic oscillator \cite{Dietl08}. 
However by computing the Hofstadter spectrum of this $\alpha$-${\cal T}_3$ semi-Dirac model for small magnetic field and various $\alpha$,
we have observed that the effective Landau levels $\epsilon_n(B)$ are well described by the approximate form $\epsilon_n \propto [(n+1/2)B]^{2/3}$, first obtained in~\cite{Dietl08}.
Interestingly, for $n\ge 3$ the numerical Landau levels $\epsilon_n(B)$ appear to be almost independent of $\alpha$.
This observation implies that the dia- to paramagnetic transition is essentially driven by the variation of the $n=0,1,2$ Landau levels.
We also stress that for this semi-Dirac model, there is no Berry phase around the semi-Dirac point. However  the Berry curvature is non-zero (for $\alpha \ne 1$) and it exhibits a double-peak structure near the semi-Dirac point.
There is still a lack of a clear physical picture of the origin of the dia to paramagnetic transition for this $\alpha$-${\cal T}_3$ semi-Dirac model.
As a final remark, we note that since the low-energy Hamiltonian (\ref{lowhamil2}) is still separable (i.e. $h^{xy}_\k=0$), the Fukuyama formula~\cite{Fukuyama71} would have given the same result.


\subsection{$\alpha$-${\cal T}_3$ {\it pseudo-bilayer} model}
\medskip

\begin{figure}
\begin{center}
\begin{tabular}{cc}
{\includegraphics[width=4.5cm]{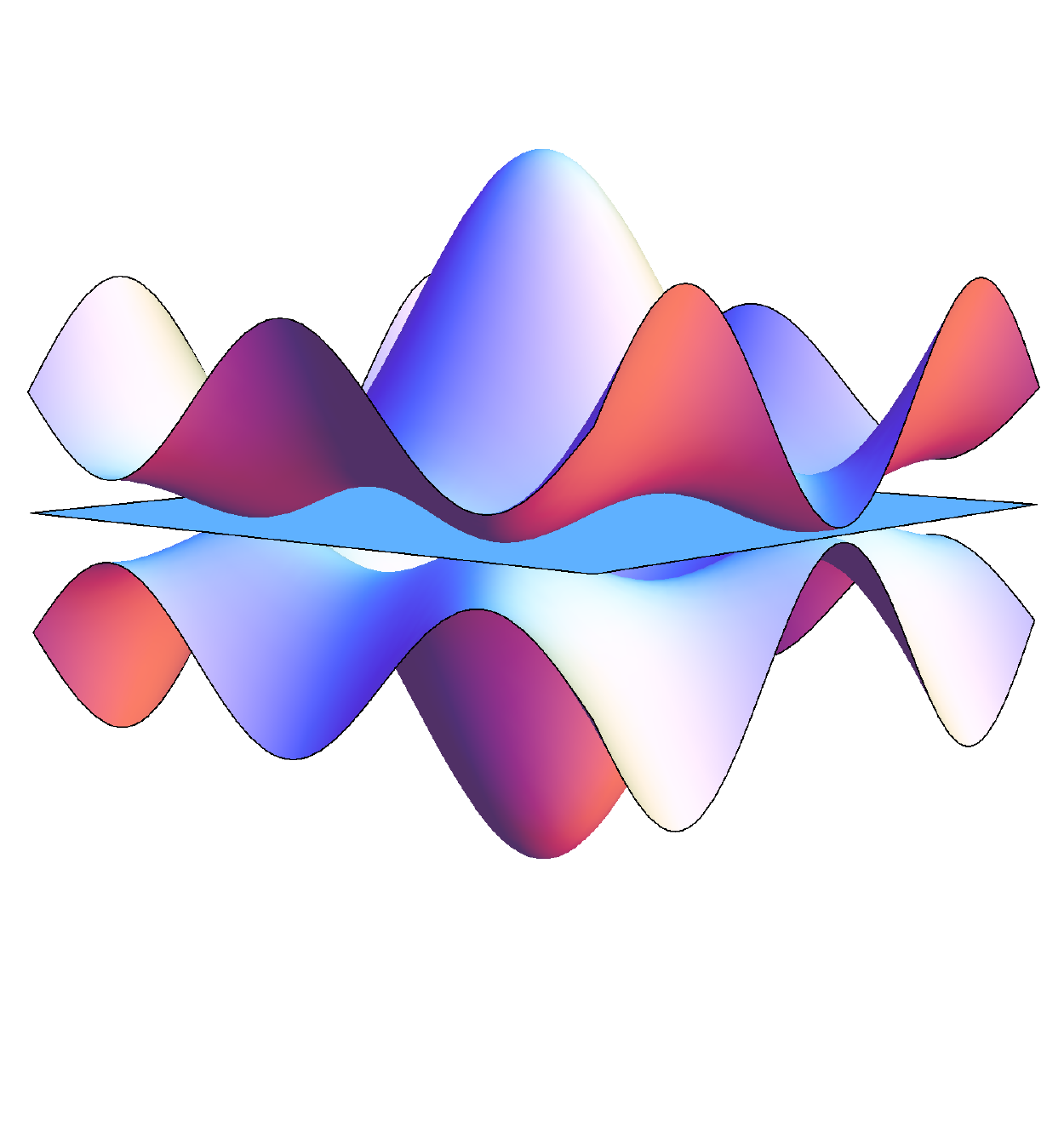}}&
{\includegraphics[width=3.5cm]{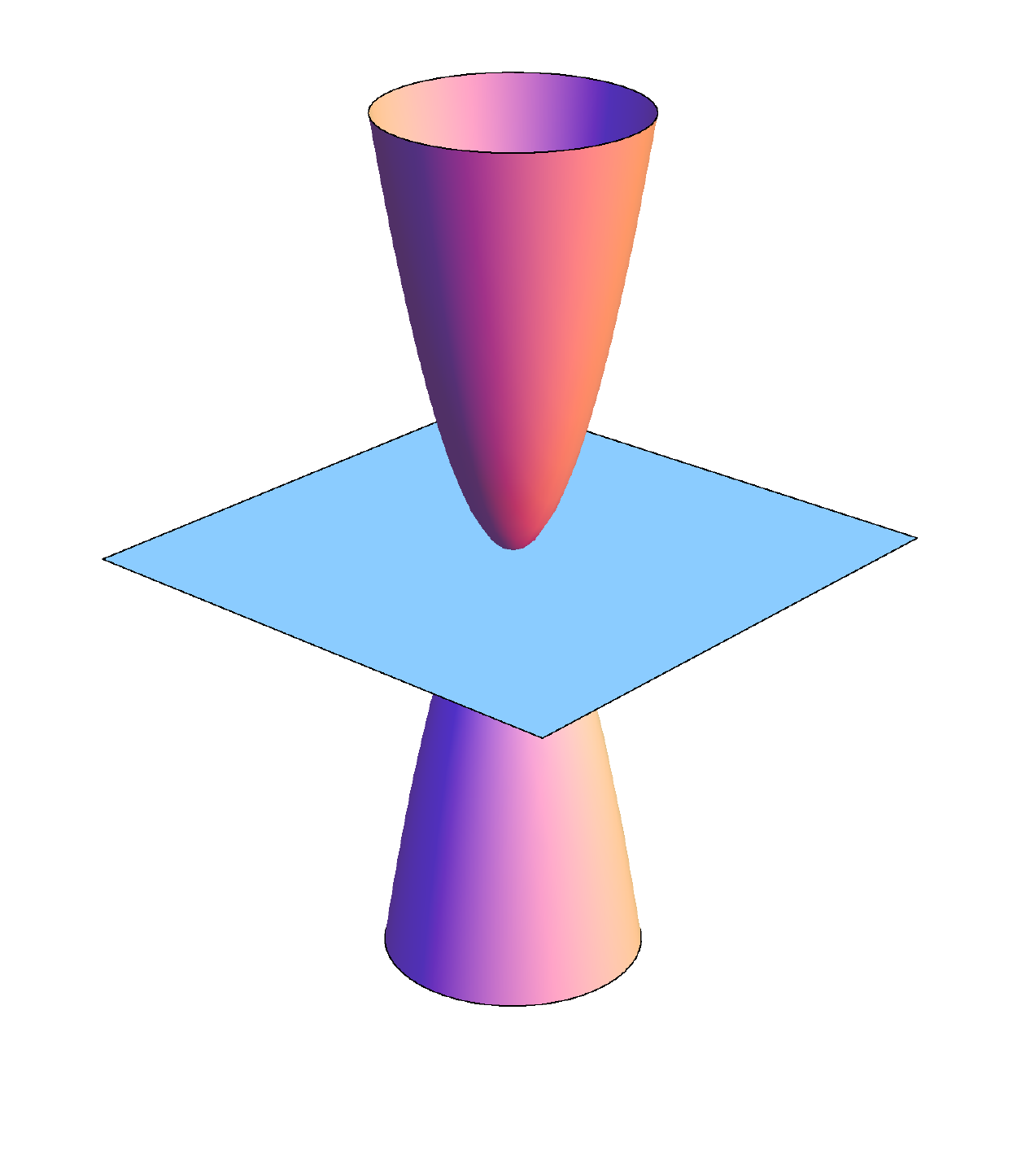}}\\
 \Large $(a)$&\Large$(b)$
 \end{tabular}
\caption{(Color online). $\alpha$-${\cal T}_3$ {\it pseudo-bilayer} model
energy bands dispersion in $\k$ space, for $\Delta=0$. (a) Full energy spectrum with quadratic band touching. (b) low-energy spectrum.}
\label{fig:t3bilayer}
\end{center}
\end{figure}

The last model we consider has parameters $t'=t$ and $t_3=1/2$ in Fig.~\ref{fig:t3bilayer}~\cite{Montambaux12}.
In that situation,
\begin{equation}
 f_\k= t(e^{-ik_ya}+2e^{ik_ya/2}\cos(\sqrt3k_xa/2))+\frac{t}{2}(e^{2ik_ya}+2e^{-ik_ya}\cos(\sqrt3k_xa) )\\
\end{equation}
For $\Delta=0$, the corresponding $\alpha$-${\cal T}_3$ energy spectrum (see Fig.~\ref{fig:t3bilayer}) exhibits quadratic band touching at $\pm {\bm K}$ similar to a bilayer graphene.
The low-energy model obtained by expanding near $\xi {\bm K}$ reads $f_{\xi {\bm K} +\k}\simeq \frac{1}{2m_*} (\xi k_x-ik_y)^2$ with the effective mass $m_*=\frac{2}{ta^2}$ and
the energy spectrum $\varepsilon_\k=\frac{\k^2}{2m_*}$. As in previous sections, to simplify the notations of most equations, we define the pseudo wavevectors $\kappa_{x,y}=\frac{ k_{x,y}}{\sqrt{2m_*}}$.

The low-energy $\alpha$-${\cal T}_3$ Hamiltonian becomes:
\be
\label{lowhamil3}
h_{\k} =
\left(
\begin{array}{ccc}
            \Delta&  c_\alpha (\xi \kappa_x-i\kappa_y)^2 & 0 \\
            c_\alpha (\xi \kappa_x+i\kappa_y)^2 & -\Delta&  s_\alpha   (\xi \kappa_x-i\kappa_y)^2\\
            0 &    s_\alpha (\xi \kappa_x+i\kappa_y)^2& \Delta \\
\end{array}
\right) \, .
\ee
This effective Hamiltonian is now quadratic in $\kappa_{x,y}$ and it is not separable (i.e. $h_\k ^{x,y}\ne 0$). As a consequence, we do not expect the Fukuyama formula to give
the correct result in that situation. Following similar steps as in previous examples we obtain
\be
\ba{lll}
\vec{f}=(\kappa_x^2-\kappa_y^2, -2\xi \kappa_x \kappa_y),&\vec{f}^x=2(\kappa_x,-\xi \kappa_y),& \vec{f}^y= 2(-\kappa_y,-\xi \kappa_x),\\
\vec{f}^{xx}=2(1,0),& \vec{f}^{yy}=2(-1,0),& \vec{f}^{xy}=2(0,-\xi),
\ea
\ee
such that by noting  $\kappa_x= \sqrt{\varepsilon} \cos{\theta}$, $\kappa_y= \sqrt{\varepsilon} \sin{\theta}$
we deduce
\begin{equation}
\label{eq:uvkbilayer}
\ba{l}
U_\k(\omega)=-16 [\frac{1}{(\omega^2-(\Delta^2+\varepsilon^2))}+\frac{\varepsilon^2}{(\omega^2-(\Delta^2+\varepsilon^2))^2}],\\
V_\k(\omega)= 16 \varepsilon^2 \left[\frac{1}{(\omega^2-(\Delta^2+\varepsilon^2))^2}[1+3\left(\frac{1-\alpha^2}{1+\alpha^2}\right)^2]
+\frac{4\varepsilon^2}{(\omega^2-(\Delta^2+\varepsilon^2))^3}\right].

%
\ea
\end{equation}
Substituting these expressions in (\ref{eq:chi2}), summing over the two valley, integrating over $\theta$ we find
\begin{equation}
\label{eq:bilayer}
\ba{ll}
\chi_\mathrm{orb}(\mu)&=\frac{\mu_0 e^2 }{12}\left(\frac{8}{\pi}\right)
\frac{\Im m}{\pi }\int_{-\infty}^{+\infty} \ud \omega \ n_\mathrm{F}(\omega)  \\
&\times \int_0 ^{\infty} \ud \varepsilon \left[\frac{\varepsilon^2}{(\omega^2-(\Delta^2+\varepsilon^2))^2}[5+12\left(\frac{1-\alpha^2}{1+\alpha^2}\right)^2]
+\frac{16\varepsilon^4}{(\omega^2-(\Delta^2+\varepsilon^2))^3}+\frac{1}{(\omega^2-(\Delta^2+\varepsilon^2))}\right].
\ea
\end{equation}
From Eqs.(\ref{appchi3},\ref{appchi5}) given in appendix B, we finally obtain:
\begin{equation}
\label{eq:mergingb}
\chi_\mathrm{orb}(\mu)=-\frac{\mu_0 e^2}{\pi m_*}
\times \left\lbrace
\ba{ll}
[\left(\frac{1-\alpha^2}{1+\alpha^2}\right)^2-\frac{3}{4}]\ln\frac{\epsilon_c}{|\mu|}-[\left(\frac{1-\alpha^2}{1+\alpha^2}\right)^2-\frac{11}{12}]  & \Delta=0,|\mu|<\epsilon_c, T=0\\
&\\
{\rm [}\left(\frac{1-\alpha^2}{1+\alpha^2}\right)^2-\frac{3}{4}{\rm ]}\ln\frac{2\epsilon_c}{\Delta}-[\left(\frac{1-\alpha^2}{1+\alpha^2}\right)^2-\frac{11}{12}]& |\mu|<\Delta, T=0
\ea
\right.
\end{equation}
where it is necessary to introduce a cutoff scale $\epsilon_c$~\cite{Safran84,Koshino11}.
As in previous examples, there is a diamagnetic to paramagnetic transition of the logarithmic term at a critical value $\alpha_c=\sqrt{4}-\sqrt{3}=0.263$, whereas
there is a paramagnetic to diamagnetic transition of the second term at $\alpha_c=\sqrt{12}-\sqrt{11}=0.147$.
For $\alpha=0$, these expressions coincide with previous results obtained in~\cite{Safran84,Koshino11}.
In ~\cite{Koshino11}, the susceptibility was derived from using the analytical form of the effective Landau levels $\epsilon_n=\epsilon_B \sqrt{n(n-1)}$
($\epsilon_B=\frac{e B}{m_*}$) that are associated to the low-energy model. A similar derivation for any $\alpha$ seems difficult to achieve even if the Landau level sequence is known.
In that perspective, we note that for the gapless $\alpha$-${\cal T}_3$ {\it pseudo-bilayer} model it is possible to derive a topological Berry phase that varies with $\alpha$;
this already gives some insight on the modified semiclassical Landau levels spectrum~\cite{Fuchs10}. However as noted in~\cite{Fuchs10}, for the bilayer system the semiclassical
Landau levels spectrum does not
fully agree with the exact quantum Landau levels spectrum and therefore we do not expect the semiclassical Landau levels spectrum to be sufficient to describe
correctly the susceptibility --especially
the singular behaviour of the susceptibility near half filling $\mu=0$.

\section{Summary}
\medskip

In this work we have studied the importance of interband effects on the orbital susceptibility of three bands $\alpha$-${\cal T}_3$ tight-binding models that were initiated in~\cite{Raoux14}.
The particularity of the $\alpha$-${\cal T}_3$ tight-binding models is that the coupling between the three energy bands (which is encoded in the wavefunctions properties) can be tuned (by a parameter $\alpha$)
without any modification of the energy spectrum.
To highlight the role of these interbands effects, we have examined the orbital susceptibility of these models.
Using the gauge-invariant perturbative formalism that we have recently developped~\cite{Raoux14b}, we obtained the generic formula of the orbital susceptibility of the $\alpha$-${\cal T}_3$ tight-binding models.
More quantitatively we have calculated the orbital susceptibility of three distinct $\alpha$-${\cal T}_3$ tight-binding models: $\alpha$-${\cal T}_3$ graphene model, the $\alpha$-${\cal T}_3$
semi-Dirac model and the $\alpha$-${\cal T}_3$ {\it pseudo-bilayer} model.
To obtain an analytical form of the susceptibility, we have only considered a low-energy Hamiltonian of these models; it correctly describes the energy spectrum
near the band touching where it is expected that interband effects are the strongest.
The main result of our study is that for each of these models, by varying the parameter $\alpha$ and thus the wavefunctions interband coupling, it is possible to drive a transition from a diamagnetic
to a paramagnetic behavior, both in absence or presence of a gap separating the dispersive bands.
In particular we emphasize that the in-gap susceptibility does not need to be diamagnetic. Moreover the existence of a finite in-gap (paramagnetic) susceptibility at $\alpha=1$ (for each model)
provides hints that some important interband effects are not encoded in the Berry curvature which vanishes for $\alpha=1$.




\appendix
\section{Determination of  $U_\k(\omega)$ and $V_\k(\omega)$}
\medskip

The aim of this section is to give the key step to find the expressions $U_\k(\omega)$ and $V_\k(\omega)$ of Eq.(\ref{eq:uvkb}).
We first slightly rewrite their definitions Eq.(\ref{eq:uvk}):
\begin{equation}
\label{appuvk}
\ba{l}
U_\k(\omega)=\mathrm{tr}\left\{(gh^{xx}gh^{yy}-gh^{xy}gh^{xy})_\k\right\},\\
V_\k(\omega)=\mathrm{tr}\left\{(gh^xgh^y[gh^y,gh^x])_\k\right\},
\ea
\end{equation}
where $g_\k(\omega)=(\omega {\rm{I}}-h_\k)^{-1}$ it the Green's function matrix associated to the zero-field Bloch Hamiltonian matrix Eq.(\ref{hamiltonian}),
 $h_\k ^{i}=\frac{\partial h_\k}{\partial {k_i}}$ and $h_\k ^{ij}=\frac{\partial^2 h_\k}{\partial {k_i}\partial{k_j}}$ with $(i,j)=(x,y)$.
The calculation of $U_\k,V_\k$ is made simple by remarking that the matrices $h,g,h^j,h^{ij}$ can each be written in terms of the following three matrices only:
\be
\ba{ccc}
S_{\pm}=\left(
\begin{array}{ccc}
            0 & \pm c_\alpha e^{-i\theta_\k}& 0 \\
            c_\alpha e^{i\theta_\k} & 0 &  \pm s_\alpha e^{-i\theta_\k} \\
            0  &  s_\alpha e^{i\theta_\k}&0 \\
\end{array}
\right) \ ;&
S_0=\left(
\begin{array}{ccc}
            1 & 0 & 0 \\
           0  & -1 & 0  \\
            0  & 0 &1 \\
\end{array}
\right),
\ea
\ee
Note that the two matrices $S_\pm$ depend on both parameter $\alpha$ and wavevector $\k$; moreover
we stress that $S_-$ is an antihermitian matrix whereas $S_{+,0}$ are hermitian.
With these definitions, it is immediate to verify the identities
\begin{equation}
\label{mat}
 \ba{l}
 h=\Delta S_0+|f_\k| S_+,\\
 h^j=u_\k ^j S_+ +i v_\k ^j S_-,\\
 h^{ij}=u_\k ^{ij} S_+ +i v_\k ^{ij} S_-,\\
 \ea
\end{equation}
with
\begin{equation}
\label{appvec1}
\ba{l}
u_\k^j=\frac{\vec{f}_\k^j \cdot \vec{f}_\k}{|f_\k|}, \ \ \ v_\k^j=\frac{(\vec{f}_\k \times \vec{f}_\k^j)}{|f_\k|}, \\
u_\k^{ij}=\frac{\vec{f}_\k^{ij} \cdot \vec{f}_\k}{|f_\k|}, \ \ \ v_\k^{ij}=\frac{(\vec{f}_\k \times \vec{f}_\k^{ij})}{|f_\k|},
\ea
\end{equation}
where we have defined the two components vectors ($\vec{x}\equiv(x_1,x_2)$)
\begin{equation}
\label{appvec}
\ba{l}
\vec{f}_\k=(\Re e f_\k, \Im m f_\k),\\
\vec{f}_\k ^i= (\frac{\partial \Re e f_\k}{\partial {k_i}},\frac{\partial \Im m f_\k}{\partial {k_i}}),\\
\vec{f}_\k ^{ij}=(\frac{\partial^2\Re e f_\k}{\partial {k_i}\partial{k_j}},\frac{\partial^2 \Im m f_\k}{\partial {k_i}\partial{k_j}}).
\ea
\end{equation}
To obtain a simple form for $g$ we remark that the matrices $S_{\pm,0}$ appear to have very peculiar properties: $S_{\pm} ^3=\pm S_{\pm}$, $S_0 ^2=I$, $[S_{\pm},S_0]_+=0$.
From these properties we deduce that $g=(\omega {\rm{I}}-h_\k)^{-1}$ can be written as $g=a_1 I +a_2 S_0+a_3 S_+ +a_4 S_+ ^2$ where $a_{1,2,3,4}$ are determined from using
the identity $(\omega-h)g={\rm I}$. We then obtain:
\begin{equation}
\label{green}
 g=g_+g_-(\omega {\rm{I}} -\Delta S_0 -|f_\k| S_+) -g_+g_-g_0 |f_\k|^2({\rm{I}}-S_+ ^2),
\end{equation}
where $g_{\pm \k}(\omega)=\frac{1}{\omega-\epsilon_{\pm,\k}}$ and $g_{0}(\omega)=\frac{1}{\omega-\epsilon_{0}}$.
Note that $g(\omega)$ has three poles corresponding to the three bands.
Substituting identities Eqs.(\ref{mat},\ref{green}) in Eq.(\ref{appuvk}), we obtain the expressions Eq.(\ref{eq:uvkb}):
\begin{equation}
\label{appuvkb}
\ba{l}
U_\k(\omega)=2g_+g_-[(u_\k^{xx}u_\k^{yy}-u_\k^{xy}u_\k^{xy})+(v_\k^{xx}v_\k^{yy}-v_\k^{xy}v_\k^{xy})]
+4g_+^2g_-^2(u_\k^{xx}u_\k^{yy}-u_\k^{xy}u_\k^{xy}) |f_\k|^2,\\
V_\k(\omega)=(u_\k^xv_\k^y-u_\k^yv_\k^x)^2\left[g_+^2g_-^2[1+3\left(\frac{1-\alpha^2}{1+\alpha^2}\right)^2]
+4g_+^3g_-^3|f_\k|^2\right].
\ea
\end{equation}
\section{Pole decomposition and integration over variables $\varepsilon$ and $\omega$}
\medskip

The generic form of the susceptibility of the three different $\alpha$-${\cal T}_3$ models is:
\begin{equation}
\label{appchi1}
\ba{ll}
 \chi_\mathrm{orb}&\propto \frac{\Im m}{\pi }\int \ud \omega \ n_\mathrm{F}(\omega)
\int_0 ^{\infty} \ud \varepsilon  \ \nu(\varepsilon)\left[\frac{A\varepsilon^4}{(\omega^2-(\Delta^2+\varepsilon^2))^3}
+\frac{B\varepsilon^2}{(\omega^2-(\Delta^2+\varepsilon^2))^2}+\frac{C}{(\omega^2-(\Delta^2+\varepsilon^2))} \right]\\
&\propto \frac{\Im m}{\pi }\int \ud \omega \ n_\mathrm{F}(\omega)
\int_0 ^{\infty} \ud \varepsilon  \ \nu(\varepsilon)
\left[A\varepsilon^4 (g_+ g_-)^3 +B\varepsilon^2 (g_+ g_-)^2 +C(g_+ g_-)\right]
\ea
\end{equation}
 where $g_s(\omega)=1/(\omega-\varepsilon_s)$ and with $\varepsilon_s =s \sqrt{\Delta^2+\varepsilon^2}$ ($s=\pm$).
 The effective density of states $\nu(\varepsilon)$ and the parameters $A,B,C$ of the three models ($\alpha$-graphene, $\alpha$-semi-Dirac, $\alpha$-bilayer) are
 summarized in table \ref{table1}.

\begin{table}[h]

\begin{center}
\begin{tabular}{cccc}
\hline \noalign{\smallskip}

   & $\alpha$-graphene& $\alpha$-semi-Dirac & $\alpha$-bilayer  \\

\noalign{\smallskip}\hline\noalign{\smallskip}

$\nu(\varepsilon)$  & $1/\varepsilon$ &$1/\sqrt{\varepsilon}$  & 1  \\

A  & 4 & 4 & 16  \\

B  & $1+3\left(\frac{1-\alpha^2}{1+\alpha^2}\right)^2$ & $1+3\left(\frac{1-\alpha^2}{1+\alpha^2}\right)^2$ & $5+12\left(\frac{1-\alpha^2}{1+\alpha^2}\right)^2$ \\

C  &  0& 0 & 1  \\

\noalign{\smallskip}\hline
\end{tabular}
\end{center}
\caption{ Effective density of states $\nu(\varepsilon)$ and parameters $A,B,C$ of the three considered models: $\alpha$-graphene, $\alpha$-semi-Dirac and $\alpha$-bilayer.
 }
\label{table1}
\end{table}

 We now describe the key steps to perform the explicit integration over variables $\varepsilon$ and $\omega$.
The first step consists to separate the poles at $\varepsilon_s$:
\begin{equation}
\label{appchi2}
\ba{l}
\chi_\mathrm{orb}\propto \sum_{s=\pm}  \frac{\Im m}{\pi }\int \ud \omega \ n_\mathrm{F}(\omega)
\int_0 ^{\infty} \ud \varepsilon  \ \nu(\varepsilon)
\left[ \frac{A}{8}\frac{\varepsilon^4}{\varepsilon_s^3} g_s ^3 +(\frac{B}{4}\frac{\varepsilon^2}{\varepsilon_s^2}-3\frac{A}{16}\frac{\varepsilon^4}{\varepsilon_s^4})g_s ^2
+(\frac{C}{2\varepsilon_s}-\frac{B}{4}\frac{\varepsilon^2}{\varepsilon_s^3} +3\frac{A}{16}\frac{\varepsilon^4}{\varepsilon_s^5}))g_s\right]
\ea
\end{equation}
The next step consists to use the identity $\frac{\Im m}{\pi } \int \ud \omega \ n_\mathrm{F}(\omega)g_s^n(\omega) =-\frac{1}{(n-1){\rm{!}}}n_\mathrm{F}^{n-1}(\varepsilon_s)$
where $n_\mathrm{F}^{(n)}$ is the $n^{th}$ derivative; we can the rewrite:
\begin{equation}
\label{appchi2}
\ba{l}
\chi_\mathrm{orb}\propto -\sum_{s=\pm}
\int_0 ^{\infty} \ud \varepsilon  \ \nu(\varepsilon)
\left[ \frac{A}{16}\frac{\varepsilon^4}{\varepsilon_s^3} n_\mathrm{F}^{''}(\varepsilon_s)
+(\frac{B}{4}\frac{\varepsilon^2}{\varepsilon_s^2}-3\frac{A}{16}\frac{\varepsilon^4}{\varepsilon_s^4})n_\mathrm{F}^{'}(\varepsilon_s)
+(\frac{C}{2\varepsilon_s}-\frac{B}{4}\frac{\varepsilon^2}{\varepsilon_s^3} +3\frac{A}{16}\frac{\varepsilon^4}{\varepsilon_s^5})n_\mathrm{F}(\varepsilon_s)\right]
\ea
\end{equation}
From this point, we separate the study of case $\Delta=0$ from case $\Delta \ne 0$.\\

{\em Gapless models, $\Delta=0$;}\\

For $\Delta=0$, since $\varepsilon_s=s \varepsilon$ we first rewrite:
\begin{equation}
\label{appchi3}
\ba{l}
\chi_\mathrm{orb}\propto -
\int_0 ^{\infty} \ud \varepsilon  \ \nu(\varepsilon)
\left[ \frac{A}{16} \varepsilon (n_\mathrm{F}^{''}(\varepsilon)-n_\mathrm{F}^{''}(-\varepsilon))
+(\frac{B}{4}-3\frac{A}{16})(n_\mathrm{F}^{'}(\varepsilon)+n_\mathrm{F}^{'}(-\varepsilon))
+(\frac{C}{2}-\frac{B}{4} +3\frac{A}{16})\frac{n_\mathrm{F}(\varepsilon)-n_\mathrm{F}(-\varepsilon)}{\varepsilon}\right]
\ea
\end{equation}
We now note that for the three considered cases, $\nu(\varepsilon)\sim \varepsilon^p$ with $-1\le p\le 0$;
this property permits an integration by part of the term proportionnal to $n_F^{''}$ and $n_F$:
\begin{equation}
\ba{l}
\int_0 ^{\infty} \ud \varepsilon \varepsilon^{p+1} (n_\mathrm{F}^{''}(\varepsilon)-n_\mathrm{F}^{''}(-\varepsilon))=
\left[ \varepsilon^{p+1} (n_\mathrm{F}^{'}(\varepsilon)+n_\mathrm{F}^{'}(-\varepsilon))\right]_0^{\infty}-(p+1)
\int_0 ^{\infty} \ud \varepsilon \varepsilon^{p} (n_\mathrm{F}^{'}(\varepsilon)+n_\mathrm{F}^{'}(-\varepsilon)),\\

\int_0 ^{\infty} \ud \varepsilon \varepsilon^{p-1} (n_\mathrm{F}(\varepsilon)-n_\mathrm{F}(-\varepsilon))=
\left[ \frac{\varepsilon^{p}}{p} (n_\mathrm{F}(\varepsilon)-n_\mathrm{F}(-\varepsilon))\right]_0^{\infty}-
\int_0 ^{\infty} \ud \varepsilon \frac{\varepsilon^{p}}{p} (n_\mathrm{F}^{'}(\varepsilon)+n_\mathrm{F}^{'}(-\varepsilon)),
\ea
\end{equation}
where the last line requires $p<0$.
Using these identities
for the three considered cases ($p=-1,-1/2,0$) and using the parameters of table \ref{table1} we deduce:
\begin{equation}
\ba{ll}
\chi_\mathrm{orb}\propto\frac{3}{2} [\left(\frac{1-\alpha^2}{1+\alpha^2}\right)^2-\frac{1}{3}]n_\mathrm{F}^{'}(0),& \alpha{\rm -graphene}\\
\chi_\mathrm{orb}\propto \frac{3}{4} [\left(\frac{1-\alpha^2}{1+\alpha^2}\right)^2-\frac{1}{2}]\int_0 ^{\infty} \ud \varepsilon
\frac{n_\mathrm{F}^{'}(\varepsilon)+n_\mathrm{F}^{'}(-\varepsilon)}{\sqrt{\varepsilon}}
& \alpha{\rm -semi-Dirac}\\
\chi_\mathrm{orb}\propto 3 \left[ [\left(\frac{1-\alpha^2}{1+\alpha^2}\right)^2-\frac{3}{4}]
\int_0 ^{\infty} \ud \varepsilon  \ \frac{n_\mathrm{F}(\varepsilon)-n_\mathrm{F}(-\varepsilon)}{\varepsilon}+
[\left(\frac{1-\alpha^2}{1+\alpha^2}\right)^2-\frac{11}{12}]
\right]& \alpha{\rm -bilayer}
\ea
\end{equation}\\

{\em Gapped models, $\Delta\ne 0$:}\\

For the gapped case, we can also perform some integration by part of terms proportionnal to $n_F^{''}$ and $n_F$.
More precisely using the identities $\frac{\partial n_F^{n}}{\partial \varepsilon}=\frac{\varepsilon}{\varepsilon_s}n_F ^{n+1}$
and $\frac{\partial \varepsilon_s}{\partial \varepsilon}=\frac{\varepsilon}{\varepsilon_s}$ we first obtain
\begin{equation}
\ba{ll}
\int_0 ^{\infty} \ud \varepsilon \frac{\varepsilon^{p+4}}{\varepsilon_s^3}n_F^{''}&=
-\int_0 ^{\infty} \ud \varepsilon \frac{\varepsilon^{p+2}}{\varepsilon_s^2}
(p+3-2\frac{\varepsilon^{2}}{\varepsilon_s^2})n_F^{'},  \\
\int_0 ^{\infty} \ud \varepsilon \frac{\varepsilon^{p+2}}{\varepsilon_s^3}n_F&=
\left[-\frac{\varepsilon^{p+1}}{\varepsilon_s} n_F\right]_0^{\infty}+\int_0 ^{\infty}
\ud \varepsilon \left[\frac{(p+1)\varepsilon^{p}}{\varepsilon_s}n_F+\frac{\varepsilon^{p+2}}{\varepsilon_s^2}n_F^{'}\right],\\
\int_0 ^{\infty} \ud \varepsilon \frac{\varepsilon^{p+4}}{\varepsilon_s^5}n_F&=
\frac{1}{3}\left[-\frac{\varepsilon^{p+3}}{\varepsilon_s^3} n_F\right]_0^{\infty} +
\frac{1}{3}\int_0 ^{\infty} \ud \varepsilon \left[\frac{(p+3)\varepsilon^{p+2}}{\varepsilon_s^3}n_F+\frac{\varepsilon^{p+4}}{\varepsilon_s^4}n_F^{'}\right],\\
&=\left[-\frac{\varepsilon^{p+3}}{\varepsilon_s^3} n_F\right]_0^{\infty}+\frac{p+3}{3}\left[-\frac{\varepsilon^{p+1}}{\varepsilon_s} n_F\right]_0^{\infty}
+\frac{1}{3}\int_0 ^{\infty} \ud \varepsilon \left[\frac{(p+3)(p+1)\varepsilon^{p}}{\varepsilon_s}n_F+
\frac{(p+3)\varepsilon^{p+2}}{\varepsilon_s^2}n_F^{'}+\frac{\varepsilon^{p+4}}{\varepsilon_s^4}n_F^{'}\right]
\ea
\end{equation}
from which we deduce:
\begin{equation}
\label{appchi4}
\ba{ll}
\chi_\mathrm{orb}\propto -\sum_{s=\pm} &\left[\frac{A}{16}\left[-\frac{\varepsilon^{p+3}}{\varepsilon_s^3} n_F\right]_0^{\infty}+[\frac{B}{4}-\frac{A}{16}(p+3)]\left[\frac{\varepsilon^{p+1}}{\varepsilon_s} n_F\right]_0^{\infty}
\right. \\
&\left. +(\frac{C}{2}-(p+1)[\frac{B}{4}-\frac{A}{16}(p+3)] )
\int_0 ^{\infty} \ud \varepsilon  \ \frac{\varepsilon^p}{\varepsilon_s}n_\mathrm{F}(\varepsilon_s)\right].
\ea
\end{equation}
For the three considered cases ($p=-1,-1/2,0$) with the parameters of table \ref{table1} we finally obtain:
\begin{equation}
\label{appchi5}
\ba{ll}
\chi_\mathrm{orb}\propto \frac{3}{4} [\left(\frac{1-\alpha^2}{1+\alpha^2}\right)^2-\frac{1}{3}]\frac{n_F(\Delta)-n_F(-\Delta)}{\Delta}& \alpha{\rm -graphene},\\
\chi_\mathrm{orb}\propto  \frac{3}{8} [\left(\frac{1-\alpha^2}{1+\alpha^2}\right)^2-\frac{1}{2}]\int_{\Delta}^{\infty}
\ud \varepsilon \frac{n_\mathrm{F}(\sqrt{\Delta^2+\varepsilon^2})-n_\mathrm{F}(-\sqrt{\Delta^2+\varepsilon^2})
}{\sqrt{\varepsilon}\sqrt{\Delta^2+\varepsilon^2}} &\alpha{\rm -semi-Dirac}\\
\chi_\mathrm{orb}\propto 3\left[[\left(\frac{1-\alpha^2}{1+\alpha^2}\right)^2-\frac{3}{4}]\int_{\Delta}^{\infty}  \ud \varepsilon
\frac{n_\mathrm{F}(\sqrt{\Delta^2+\varepsilon})-n_\mathrm{F}(-\sqrt{\Delta^2+\varepsilon})
}{\sqrt{\Delta^2+\varepsilon^2}}-[\left(\frac{1-\alpha^2}{1+\alpha^2}\right)^2-\frac{11}{12}]\right], &\alpha{\rm -bilayer}\\
\ea
\end{equation}

\newpage



\end{document}